\documentstyle[epsf,twoside,fleqn,espcrc2]{article}
\title{Finite-temperature chiral transitions in QCD with the Wilson quark
action
\thanks{
{Talk presented by S.\ Kaya
at Lattice 97, Edinburgh, Scotland, July 22-26, 1997.
}}}

\author{
S.~Aoki\rlap,\address{Institute of Physics, University of Tsukuba, 
Ibaraki 305, Japan}
Y.~Iwasaki\rlap,$^{\rm a,}$\address{Center for Computational Physics, 
University of Tsukuba, Ibaraki 305, Japan}
K.~Kanaya\rlap,$\mbox{}^{\rm a,b}$
S.~Kaya\rlap,\address{Institute of Particle and Nuclear Studies,
High Energy Accelerator Research Organization(KEK), Tsukuba, Ibaraki 305, Japan}
A.~Ukawa\rlap,$\mbox{}^{\rm a}$
and
T.~Yoshi\'e$\mbox{}^{\rm a,b}$
}

\begin{document}
\renewcommand{\textfraction}{0.1}
\renewcommand{\topfraction}{0.9}
\begin{abstract}
We investigate the finite-temperature phase structure 
and the scaling of the chiral condensate
in lattice QCD with two degenerate light quarks,
using a renormalization group improved gauge action
and the Wilson quark action.
We obtain a phase diagram which is consistent with 
that from the parity-flavor breaking scenario. 
The scaling relation for the chiral condensate
assuming the critical exponents and the scaling function of
the three dimensional O(4) model is remarkably satisfied for
a wide range of parameters.
This 
indicates that the chiral transition in two flavor QCD is of second order 
in the continuum limit.

\end{abstract}

\maketitle
\setcounter{footnote}{0}
\section{Introduction}
As a step toward the clarification of the finite-temperature QCD transition, 
it is important to investigate the nature of the transition
on the lattice with two degenerate light quarks.
In a previous study\cite{Prev}, using the  Wilson quark action and  
a renormalization group (RG) improved gauge action
\begin{equation}
S_g^{R} = {\beta \over 6}\left(c_0 \sum W_{1\times 1}
               + c_1 \sum W_{1\times 2}\right),
\label{eqn:Sg}
\end{equation}
with $c_0=1-8c_1$ and $c_1=-0.331$~\cite{IM},  
we studied
the nature of the phase transition and 
the scaling behavior of the chiral condensate at
$\beta > \beta_{\rm ct}$ $(\approx 1.35)$ on an $N_t=4$ lattice,
where $\beta_{\rm ct}$ is the value of $\beta$
at the chiral transition point.
In the present work, we extend the study to 
$\beta \leq \beta_{\rm ct}$ 
performing simulations at $\beta=1.1$, 1.2, and 1.35
on an $8^3 \times 4$ lattice.
We also determine the phase structure of the chiral limit, in particular,
for $\beta \leq \beta_{\rm ct}$ at $N_t=4$
on a $(\beta,K)$ plane.

\section{Phase diagram}

\begin{figure}[tb]
\begin{center}
\leavevmode
\epsfxsize=6.2cm
\epsfbox{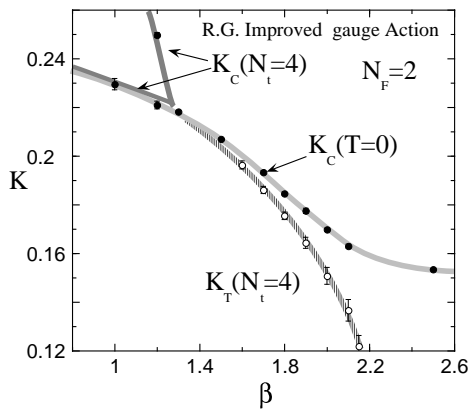}
\end{center}
\vspace{-1.5cm}
\caption{Phase diagram for $N_F=2$ with the RG improved gauge action
and the Wilson quark action.} 
\label{fig:dig}
\vspace{-0.5cm}
\end{figure}

Fig.\ref{fig:dig} shows our result for 
the phase diagram on a $(\beta,K)$ plane.
In the previous work\cite{Prev},
the line of the zero-temperature chiral limit $K_c(T=0)$ 
defined by the vanishing point of the pion mass was 
determined on an $8^4$ lattice,
and the finite-temperature transition/crossover line $K_t$ 
for $N_t=4$ was determined on an $8^3\times4$ lattice. 
The $K_t$ line crosses the $K_c(T=0)$ line
at $\beta_{\rm ct}\approx 1.35$.

\begin{figure}[tb]
\vspace{-0.4cm}
\begin{center}
\leavevmode
\epsfxsize=5.8cm
\epsfbox{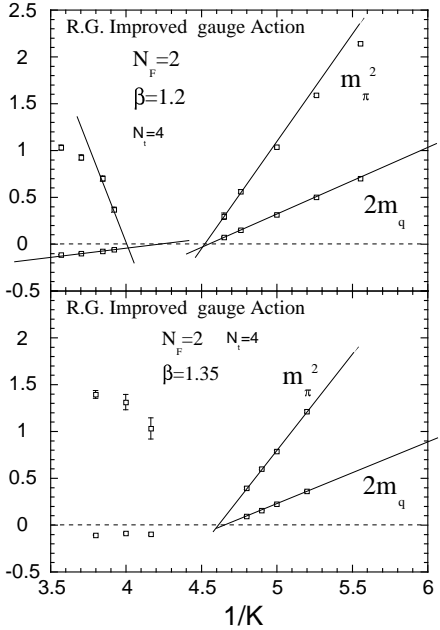}
\end{center}
\vspace{-1.4cm}
\caption{$m_{\pi}^2$ and $2m_q$ vs.\ 1/K for $N_F=2$ 
obtained at $\beta=1.2$ and 1.35 on an $8^3\times 4$ lattice.}
\label{fig:mass}
\vspace{-0.5cm}
\end{figure}

In this work, we estimate the location of the finite-temperature 
chiral limit $K_c(N_t=4)$ for $N_t=4$,
defined by the vanishing point of the pion screening mass $m_{\pi}$.
In Fig.\ref{fig:mass},
we plot $m_{\pi}^2$ and the quark mass $m_q$ for $N_t=4$ 
at fixed $\beta$ as a function of $1/K$,
where $m_q$ is defined by an axial Ward identity\cite{Bochicchio,ourMq}.
In the upper figure for $\beta=1.2$, 
there are two chiral limits. 
On the other hand, at $\beta=1.35 \approx \beta_{\rm ct}$ in the lower figure,
we cannot find a clear gap between the two chiral limits.
These results imply 
that the line $K_c(N_t=4)$ for the chiral limit
turns back towards strong coupling around $\beta_{\rm ct}$, 
forming a cusp, as shown in Fig.\ref{fig:dig}.
This structure is consistent with that expected from
the parity-flavor breaking scenario\cite{Aoki}, which
was confirmed for the standard one plaquette gauge action and 
the Wilson quark action\cite{AUU}. 

\section{Scaling analysis}

\begin{figure}[tb]
\begin{center}
\leavevmode
\epsfxsize=6.0cm
\epsfbox{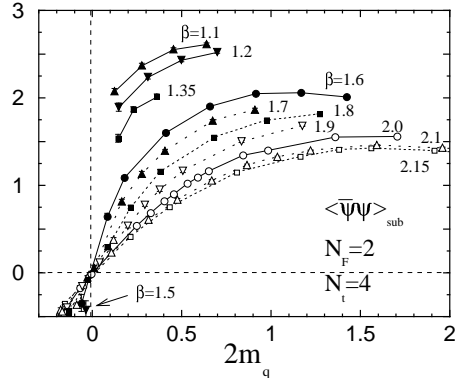}
\end{center}
\vspace{-1.5cm}
\caption{Subtracted chiral condensate 
$\langle \bar{\Psi} \Psi \rangle_{\rm sub}$
as a function of $2m_qa$.}
\label{fig:psi}
\vspace{-0.5cm}
\end{figure}

The magnetization $M$ near the second order transition
is expected to be described by a 
single scaling function;
\begin{equation}
M / h^{1/\delta} = f(t/h^{1/(\beta\delta)}),
\label{eq:universality}
\end{equation}
with $h$ the external magnetic field 
and $t=[T-T_c(h\!=\!0)]/T_c(h\!=\!0)$ the reduced temperature.

According to the universality argument,
when the chiral transition is of second order, two flavor QCD
belongs to the same universality class
as the three dimensional O(4) spin model\cite{Wilczek},
and the chiral condensate
should satisfy the scaling relation (\ref{eq:universality})
with the O(4) exponents and the O(4) scaling function.
We identify $h=2m_q a$, $t=\beta\!-\!\beta_{\rm ct}$,
and $M=\langle \bar{\Psi} \Psi \rangle_{\rm sub}$, 
where $M$ is 
a subtracted chiral condensate 
defined through an axial Ward identity\cite{Bochicchio},
\begin{equation}
\langle \bar{\Psi} \Psi \rangle_{\rm sub}
= 2 m_q a (2K)^2 \sum_x \langle \pi(x) \pi(0) \rangle.
\label{eq:PBPsub}
\end{equation}
Our results for 
$\langle \bar{\Psi} \Psi \rangle_{\rm sub}$ 
are shown in Fig.\ref{fig:psi}.

\begin{figure}[tb]
\vspace{-0.2cm}
\begin{center}
\leavevmode
\epsfxsize=6cm
\epsfbox{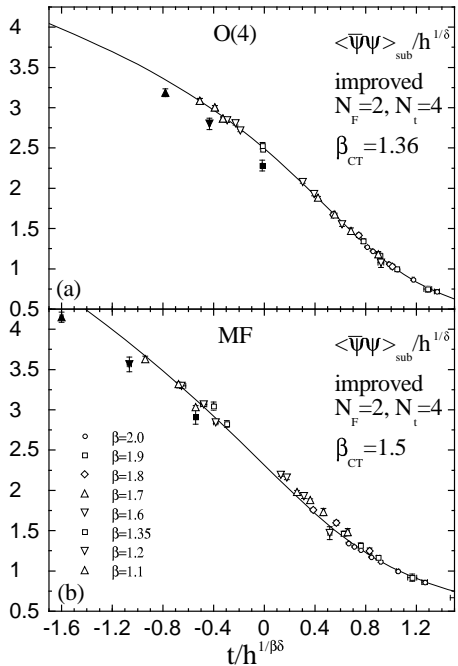}
\end{center}
\vspace{-1.5cm}
\caption{Fits to the scaling function with (a) O(4) and (b)
MF exponents. 
Solid curves are scaling functions obtained in an O(4)
spin model\protect\cite{Toussaint} and by a MF calculation, respectively.}
\label{fig:scaling}
\vspace{-0.5cm}
\end{figure}

We make a fit to the scaling function
of the O(4) model
by adjusting $\beta_{\rm ct}$ and the scales for $t$ and $h$,
with the exponents fixed to the O(4) values\cite{KanayaKaya},
including all the data in the range $0 < 2m_q a < 0.8$ and $\beta \leq 2.0$
shown in Fig.\ref{fig:psi}.
We note that the data for the lightest quark mass ($m_q \!=\! 0.06$-0.07)
at $\beta\!=\! 1.1$, 1.2, and 1.35 are slightly off the scaling curve.
Therefore, we make a fit excluding these three points.
The result of the fit with $\chi^2/df = 0.72$ 
is shown in Fig.\ref{fig:scaling}(a).
The result shows that scaling works well in the $t \leq 0$ region 
as well as the $t>0$ region,
except the three lightest quark mass data at $t < 0$,
indicated by the filled symbols.
The adjusted $\beta_{\rm ct}=1.36(1)$ is consistent with 
$\beta_{\rm ct}=1.35(1)$
obtained by the previous fit using only the $t >0$ data\cite{Prev}.

One possible origin of the deviation from the scaling curve 
of the lightest quark mass data at $t \leq 0$
is a finite size effect,
because, in the confining phase, 
we expect that finite size effects are severe at small quark mass.
To explore this possibility,
we are performing simulations on $L^3\times 4$ ($L=12$ and 16) lattices
for the lightest quark mass data at $\beta=1.1,1.2$ and 1.35.
The statistics we have accumulated so far 
(about 50 configurations each) 
is not yet high enough to obtain a definite conclusion 
about the systematic size dependence 
of the deviation from the scaling curve.

Another possible origin for the deviation
is the explicit chiral breaking due to the Wilson term,
which is expected to be large at small $\beta$.
It is plausible that this effect becomes visible as deviation from
scaling, when the explicit chiral violation due to the Wilson term
becomes larger than that from the quark mass.
The fact that we observe the deviation from the scaling at the
lightest quarks mass for small $\beta$ region is consistent
with this interpretation.
To confirm this interpretation it is necessary to make a simulation
at larger $N_t$ to reach a larger $\beta$ region, or use an improved
quark action.

We also study the possibility of the MF scaling\cite{KocicKogut}.
We perform the scaling test using MF exponents and the MF scaling function.
Because $m_{\pi}$ at $\beta$=1.5 does not vanish in the chiral limit 
(cf. Fig.3 in Ref.\cite{Prev}),
we restrict $\beta_{\rm ct} \leq 1.5$. We then obtain the best fit 
shown in Fig.\ref{fig:scaling}(b)
with $\chi^2/df = 6.2$, 
to be compared with $\chi^2/df=0.72$ obtained for the O(4) case
[Fig.\ref{fig:scaling}(a)]. 
We exclude the MF scaling, as the data is much more 
scattered than in the O(4) case.

The success of the scaling with the O(4) exponents,
albeit with the three lightest quark data at $t <0$ excluded,
indicates that the chiral transition for two flavor
QCD is of second order in the continuum limit.

\vspace{2mm}

Numerical simulations are performed with Fujitsu VPP500/30
at the University of Tsukuba.
This work is in part supported by 
the Grants-in-Aid of Ministry of Education
(Nos.\ 09246206, 09304029, 08640349, 08640350).

\end{document}